%% file: arxiv.tex
\PassOptionsToPackage{unicode}{hyperref}
\PassOptionsToPackage{hyphens}{url}
\PassOptionsToPackage{dvipsnames,svgnames,x11names}{xcolor}
\documentclass[
  12pt]{article}

\usepackage{amsmath,amssymb}
\usepackage{iftex}
\ifPDFTeX
  \usepackage[T1]{fontenc}
  \usepackage[utf8]{inputenc}
  \usepackage{textcomp} 
\else 
  \usepackage{unicode-math}
  \defaultfontfeatures{Scale=MatchLowercase}
  \defaultfontfeatures[\rmfamily]{Ligatures=TeX,Scale=1}
\fi
\usepackage{lmodern}
\ifPDFTeX\else  
\fi
\IfFileExists{upquote.sty}{\usepackage{upquote}}{}
\IfFileExists{microtype.sty}{
  \usepackage[]{microtype}
  \UseMicrotypeSet[protrusion]{basicmath} 
}{}
\makeatletter
\@ifundefined{KOMAClassName}{
  \IfFileExists{parskip.sty}{%
    \usepackage{parskip}
  }{
    \setlength{\parindent}{0pt}
    \setlength{\parskip}{6pt plus 2pt minus 1pt}}
}{
  \KOMAoptions{parskip=half}}
\makeatother
\usepackage{xcolor}
\makeatletter
\ifx\paragraph\undefined\else
  \let\oldparagraph\paragraph
  \renewcommand{\paragraph}{
    \@ifstar
      \xxxParagraphStar
      \xxxParagraphNoStar
  }
  \newcommand{\xxxParagraphStar}[1]{\oldparagraph*{#1}\mbox{}}
  \newcommand{\xxxParagraphNoStar}[1]{\oldparagraph{#1}\mbox{}}
\fi
\ifx\subparagraph\undefined\else
  \let\oldsubparagraph\subparagraph
  \renewcommand{\subparagraph}{
    \@ifstar
      \xxxSubParagraphStar
      \xxxSubParagraphNoStar
  }
  \newcommand{\xxxSubParagraphStar}[1]{\oldsubparagraph*{#1}\mbox{}}
  \newcommand{\xxxSubParagraphNoStar}[1]{\oldsubparagraph{#1}\mbox{}}
\fi
\makeatother

\usepackage{longtable,booktabs,array}
\usepackage{calc} 
\usepackage{etoolbox}
\makeatletter
\patchcmd\longtable{\par}{\if@noskipsec\mbox{}\fi\par}{}{}
\makeatother
\IfFileExists{footnotehyper.sty}{\usepackage{footnotehyper}}{\usepackage{footnote}}
\makesavenoteenv{longtable}
\usepackage{graphicx}
\makeatletter
\def\maxwidth{\ifdim\Gin@nat@width>\linewidth\linewidth\else\Gin@nat@width\fi}
\def\maxheight{\ifdim\Gin@nat@height>\textheight\textheight\else\Gin@nat@height\fi}
\makeatother
\setkeys{Gin}{width=\maxwidth,height=\maxheight,keepaspectratio}
\makeatletter
\def\fps@figure{htbp}
\makeatother

\makeatletter
\@ifpackageloaded{caption}{}{\usepackage{caption}}
\AtBeginDocument{%
\ifdefined\contentsname
  \renewcommand*\contentsname{Table of contents}
\else
  \newcommand\contentsname{Table of contents}
\fi
\ifdefined\listfigurename
  \renewcommand*\listfigurename{List of Figures}
\else
  \newcommand\listfigurename{List of Figures}
\fi
\ifdefined\listtablename
  \renewcommand*\listtablename{List of Tables}
\else
  \newcommand\listtablename{List of Tables}
\fi
\ifdefined\figurename
  \renewcommand*\figurename{Figure}
\else
  \newcommand\figurename{Figure}
\fi
\ifdefined\tablename
  \renewcommand*\tablename{Table}
\else
  \newcommand\tablename{Table}
\fi
}
\@ifpackageloaded{float}{}{\usepackage{float}}
\floatstyle{ruled}
\@ifundefined{c@chapter}{\newfloat{codelisting}{h}{lop}}{\newfloat{codelisting}{h}{lop}[chapter]}
\floatname{codelisting}{Listing}

\makeatother
\makeatletter
\@ifpackageloaded{caption}{}{\usepackage{caption}}
\@ifpackageloaded{subcaption}{}{\usepackage{subcaption}}
\makeatother

\ifLuaTeX
  \usepackage{selnolig}  
\fi
\usepackage[]{natbib}
\usepackage{bookmark}

\IfFileExists{xurl.sty}{\usepackage{xurl}}{} 
\hypersetup{
  pdftitle={Title},
  pdfauthor={Author 1; Author 2},
  pdfkeywords={3 to 6 keywords, that do not appear in the title},
  colorlinks=true,
  linkcolor={blue},
  filecolor={Maroon},
  citecolor={Blue},
  urlcolor={Blue},
  pdfcreator={LaTeX via pandoc}}

\newcommand{\anon}{1}

\usepackage{zhzhao}

\begin{document}

\def\spacingset#1{\renewcommand{\baselinestretch}%
{#1}\small\normalsize} \spacingset{1}


\if1\anon
{
  \title{\bf Nonparametric $f$-Modeling for Empirical Bayes Inference with Unequal and Unknown Variances}
  \author{Zhigen Zhao\thanks{Zhao is partially supported by NSF grant DMS-2311216.
    }\hspace{.2cm}\\
    Department of Statistics, Operations, and Data Science, Temple University\\
    Philadelphia, PA USA\\
    and \\
    Shonosuke Sugaasawa\thanks{Sugasawa is partially supported by JSPS KAKENHI Grant Numbers 24K21420 and 25H00546.} \\
    Faculty of Economics, Keio University, Tokyo, Japan}
  \maketitle
} \fi

\if0\anon
{
  \bigskip
  \bigskip
  \bigskip
  \begin{center}
    {\LARGE\bf Nonparametric $f$-Modeling for Empirical Bayes Inference with Unequal and Unknown Variances}
\end{center}
  \medskip
} \fi

\bigskip
\begin{abstract}
Empirical Bayes methods are widely used for large-scale inference, yet most classical approaches assume homoscedastic observations and focus primarily on posterior mean estimation. We develop a nonparametric empirical Bayes framework for the heteroscedastic normal means problem with unequal and unknown variances. Our first contribution is a generalized Tweedie-type identity that expresses the Bayes estimator entirely in terms of the joint marginal density of the observed statistics $(x_i, s_i^2)$ and its partial derivatives, extending the classical Tweedie’s formula to settings with heterogeneous and unknown variances. Our second contribution is to introduce a moment-generating-function representation that enables recovery of the full posterior distribution within the $f$-modeling paradigm without specifying or estimating the prior distribution. The resulting method provides a unified framework for point estimation, uncertainty quantification, and hypothesis testing while accommodating arbitrary dependence between means and variances. Simulation studies and real-data analysis demonstrate that the proposed approach achieves accurate shrinkage estimation and reliable posterior inference in heterogeneous data environments.
\end{abstract}

\noindent%
\vfill

\newpage
\spacingset{1.8} 

\section{Introduction}

\input{introduction.tex}

\section{Model with Unequal and Unknown Variances}\label{sec:model}
\input{model.tex}

\section{Tweedie's formula assuming unknown and unequal variances }\label{sec:estimation}
\input{estimation.tex}
\section{Posterior Distribution Approximation}\label{sec:interval}
\input{posterior.tex}

\section{Simulation Studies}\label{sec:simulation}
\input{simulation.tex}

\section{Data Analysis}\label{sec:realdata}
\input{realdata.tex}

\section{Concluding Remarks}\label{sec:conclusion}
\input{conclude.tex}

\label{bibtex}
\bibliography{zhaozhg.bib}

\end{document}

%% file: introduction.tex
Empirical Bayes methods, introduced by Robbins (1956), have since been developed along multiple directions. Notable developments include the seminal shrinkage estimator of \citet{James:Stein:1961}, subsequent empirical Bayes formulations by \citet{Efron:Morris:1971, Efron:Morris:1972, Efron:Morris:1973, laird1978nonparametric}, and more recent advances in large-scale inference and nonparametric empirical Bayes modeling \citep{Efron:2011, Efron:2010b, zhang1997empirical,Jiang:Zhang:2009, Brown:Greenshtein:2009, Hwang:Qiu:Zhao:2009, Zhao:Hwang:2012, gu2017empirical}. EB methods provide a flexible framework for large-scale inference by borrowing strength across many related estimation problems. They have been widely applied in genomics, neuroimaging, environmental science, and machine learning, where large numbers of related parameters must be estimated adaptively \citep{schafer2005empirical, fortin2018harmonization, gribov2020empirical} and small area estimation \citep{you2006small, sugasawa2017bayesian}.

In the classical normal means model, one observes independent data
\[
x_i \mid \theta_i \overset{\text{ind}}{\sim} N(\theta_i,\sigma^2), \quad i=1,\ldots,n,
\]
where the parameters $\theta_i$ are assumed to arise from an unknown prior distribution. A fundamental result in this setting is \emph{Tweedie’s formula}, which shows that the Bayes estimator can be expressed entirely in terms of the marginal density of the observations and its derivative \citep{Efron:2010b, efron2024empirical}. This identity has played a central role in modern empirical Bayes methodology, enabling nonparametric estimation of posterior mean and variance through estimation of the marginal density of the data. 

However, many modern applications involve substantial heterogeneity in observational variances. In large-scale experiments such as genomics, meta-analysis, and high-throughput screening studies, both the effect sizes and their variances may vary across units and are typically unknown. A common modeling framework for such settings assumes
\[
x_i \mid \theta_i,\sigma_i^2 \overset{\text{ind}}{\sim} N(\theta_i,\sigma_i^2),
\qquad
s_i^2 \mid \sigma_i^2 \overset{\text{ind}}{\sim} \sigma_i^2 \chi_k^2/k,
\]
where $s_i^2$ is an estimate of the unknown variance $\sigma_i^2$ obtained from auxiliary data or replication. Empirical Bayes inference in this heteroscedastic setting is substantially more challenging. Existing approaches often rely on parametric assumptions for the prior distributions \citep{Smyth:2004, Cui:Hwang:Qiu:Blades:Churchill:2005, Hwang:Qiu:Zhao:2009} or focus primarily on posterior mean estimation \citep{Zhao:2010}. In particular, an analogue of Tweedie’s formula that accommodates unknown and heterogeneous variances has not been established.

One of the primary contributions of this paper is to derive such an identity. We show that in the heteroscedastic normal means model with unknown variances, the Bayes estimator admits an explicit representation in terms of the joint marginal density of the observed statistics $(x_i,s_i^2)$'s and its partial derivatives. Specifically, if $f(x_i,s_i^2)$ denotes the marginal density of $(x_i,s_i^2)$, we show that the Bayes estimator can be written as
\[
\hat{\theta}_{i, Bayes}=
x+
\frac{k s_i^2 f_{x_i}(x_i,s_i^2)}
{(k-2)f(x_i,s_i^2)-2s_i^2 f_{s_i^2}(x_i,s_i^2)},
\]
where $f_{x_i}$ and $f_{s_i^2}$ denote partial derivatives of the marginal density with respect to $x_i$ and $s_i^2$, respectively. This identity can be viewed as a heteroscedastic extension of Tweedie’s formula and forms the theoretical foundation of our empirical Bayes methodology.

A classical strength of the $f$-modeling paradigm is that posterior mean can be expressed directly in terms of the marginal distribution of the data, thereby avoiding explicit estimation of the prior distribution. However, extending it to recover full posterior distributions remains technically challenging \citep{Efron:2010b, Efron:2011, efron2024empirical}. As a result, traditional $f$-modeling approaches do not directly support posterior uncertainty quantification, credible intervals, or Bayesian hypothesis testing. Meanwhile, fully nonparametric $g$-modeling typically relies on nonparametric maximum likelihood estimation of the prior distribution, which can be computationally demanding and may complicate posterior inference in practice \citep{zhang1997empirical, gu2017empirical}.

In this paper, we further propose a new nonparametric empirical Bayes framework for posterior inference under heteroscedasticity where the joint prior distribution of the means and variances is completely unspecified. Our approach builds upon a bivariate $f$-modeling formulation for the joint marginal distribution of $(x_i,s_i^2)$'s and introduces a moment generating function (MGF) representation to characterize the posterior distribution of $(\theta_i,\sigma_i^2)$. The MGF representation has several distinctive advantages. It uniquely determines the underlying distribution and provides a mechanism to recover the entire posterior distribution of $(\theta_i,\sigma_i^2)$, enabling coherent inference for point estimation, interval estimation, and multiple testing within a unified framework.

In summary, the proposed approach extends classical f-modeling empirical Bayes methodology in two important directions. First, it generalizes Tweedie-type identities to the heteroscedastic setting with unknown and unequal variances. Second, it provides a mechanism for recovering the full posterior distribution, rather than only posterior means, within the $f$-modeling framework. By providing access to the entire posterior distribution, the moment-generating-function-based empirical Bayes formulation offers a theoretically grounded and practically robust foundation for modern large-scale inference in heterogeneous data environments.

The remainder of the paper is organized as follows. Section~\ref{sec:model} presents the model formulation and derives the posterior MGF. Section~\ref{sec:estimation} develops the nonparametric empirical Bayes estimator of the mean and establishes a regret bound. Section~\ref{sec:interval} introduces the estimation of the posterior distribution of $\theta$ and a corresponding nonparametric EB confidence interval. Section~\ref{sec:testing} extends the framework to hypothesis testing. Section~\ref{sec:simulation} reports numerical studies evaluating the proposed methods. Section \ref{sec:realdata} reports the result from the baseball player data analysis. Technical proofs are collected in the Supplementary Material. The R codes to replicate the numerical results in this paper are available at GitHub repository (\url{https://github.com/sshonosuke/EBNF}).

%% file: model.tex
Suppose we have independent observation $(x_i, s_i^2)$'s which satisfy the following distribution:
\begin{eqnarray}\label{eqn:model}
\left\{ \begin{array}{c} 
x_i|\theta_i, \sigma_i^2 \overset{\text{ind}}{\sim}  N(\theta_i, \sigma_i^2), \\
s_i^2|\sigma_i^2 \overset{\text{ind}}{\sim} k^{-1}\sigma_i^2\chi_k^2.
\end{array}
\right.
\end{eqnarray}
Assume the following prior distribution 
\begin{equation}\label{eqn:prior}
(\theta_i, \sigma_i^2) \overset{\text{ind}}{\sim} g(\theta_i, \sigma_i^2).
\end{equation}
Unlike methods in the existing literature, the joint prior distribution $g$ is chosen arbitrary, allowing any dependence between $\theta_i$ and $\sigma_i^2$. In this paper, we provide inferential methods based on the full posterior distribution of $(\theta_i, \sigma_i^2)$ given the data without estimating the prior distribution.

To simplify the notation, the subscript "$i$" will be dropped when there is no ambiguity. It is easily seen that the joint likelihood function of $(x, s^2)$ is 
\begin{eqnarray}\label{eqn:likelihood}
&&p(x, s^2|\theta, \sigma^2) 
= C_k (s^2)^{k/2-1}\left(\sigma^2\right)^{-(k+1)/2}\exp\left(-\frac{(x-\theta)^2+ks^2}{2\sigma^2}\right),
\end{eqnarray}
where $C_k=k^{k/2}/\sqrt{2\pi}\Gamma(k/2)2^{k/2}$ is a constant depending on the degrees of freedom $k$ only. The joint marginal distribution of $(x,s^2)$ is 
\begin{equation}\label{eqn:marginal:pdf}
f(x, s^2) = \iint f(x,s^2|\theta,\sigma^2)g(\theta, \sigma^2)d\theta d\sigma^2.
\end{equation}
The posterior distribution of $(\theta, \sigma^2)$ is
\[
p(\theta, \sigma^2|\mathcal{D}) = \frac{p(x,s^2|\theta, \sigma^2)g(\theta, \sigma^2)}{f(x,s^2)},
\]
where $\mathcal{D}$ represents the observed data.

We reparameterize the parameter $(\theta,\sigma^2)$ by considering two random variables $U=\theta/\sigma^2|\mathcal{D}$ and $V=1/\sigma^2|\mathcal{D}$. In the next theorem, we develop the moment generating function of $U$ and $V$.

\begin{theorem}\label{thm:mgf:individual}
Assume Model~\eqref{eqn:model} and an arbitrary prior distribution~\eqref{eqn:prior}. 
For a given $(x,s^2)$ with $s^2>0$,

\begin{enumerate}
\item for all $t$ in a neighborhood of $0$ such that
\[
s^2-\frac{2tx+t^2}{k}>0,
\]
the moment generating function of $U$ is
\[
M_U(t)
= \left( \frac{s^2}{s^2-2tx/k-t^2/k}\right)^{k/2-1}
\frac{ f(x+t,\, s^2-(2tx+t^2)/k)}{f(x, s^2)};
\]

\item for all $t$ in a neighborhood of $0$ such that
\[
s^2-\frac{2t}{k}>0,
\]
the moment generating function of $V$ is
\[
M_V(t)
= \left( \frac{s^2}{s^2-2t/k}\right)^{k/2-1}
\frac{ f(x,\, s^2-2t/k)}{f(x, s^2)};
\]

\item for all $(t_1,t_2)$ in a neighborhood of $(0,0)$ such that
\[
s^2-\frac{2t_1x+t_1^2+2t_2}{k}>0,
\]
the joint moment generating function of $(U,V)$ is
\[
M_{U,V}(t_1,t_2)
= \left(\frac{s^2}{s^2-(2t_1x+t_1^2+2t_2)/k}\right)^{k/2-1}
\frac{f(x+t_1,\, s^2-(2t_1x+t_1^2+2t_2)/k)}{f(x,s^2)}.
\]
\end{enumerate}
\end{theorem}

Theorem~\ref{thm:mgf:individual} shows that the moment generating function of $(U,V)$ is completely determined by the joint marginal density of $(x,s^2)$, with the prior distribution influencing the representation only implicitly through this marginal relationship. 
The representation holds locally in neighborhoods of $0$ (or $(0,0)$ in the bivariate case), where the conditional MGFs are finite and the shifted variance arguments remain positive. 
These local conditions are sufficient for recovering the posterior distribution. 
Consequently, empirical Bayes inference can be formulated entirely in terms of estimating a smooth joint marginal density and its derivatives, without direct estimation of the prior distribution.

This perspective differs fundamentally from conventional $f$-modeling approaches, which typically focus on posterior means or variances as primary inferential targets. 
While such moment-based summaries are convenient, they may fail to capture important features of the posterior distribution in complex settings, such as heavy tails or multimodality. 
By working directly with the moment generating function, our framework provides a unified foundation for point estimation, hypothesis testing, and confidence interval construction, developed in subsequent sections.

%% file: estimation.tex
\subsection{ Bayes estimator under weighted loss}
We consider the estimation of the parameter $\theta$ with the weighted loss function
\begin{equation}\label{eqn:loss}
L(\hat{\theta}, \theta) =\frac{1}{\sigma^2}(\hat{\theta} - \theta)^2.
\end{equation}
Simple calculation shows that the Bayes estimator is 
\[
\hat{\theta}_{B} = \frac{ \mathbb{E}(\theta/\sigma^2|x,s^2)}{\mathbb{E}(1/\sigma^2|x,s^2)} = \frac{\mathbb{E}(U)}{\mathbb{E}(V)}.
\]
We next derive an explicit expression for the Bayes estimator. To simplify the notation, we use \( f_x(x, s^2) \) and \( f_{s^2}(x, s^2) \) to denote the partial derivatives $\partial f(x, s^2)/\partial x$ and $\partial f(x, s^2)/\partial s^2$, respectively.

\begin{theorem}\label{thm:Bayes}
    Assume the model (\ref{eqn:model}), any arbitrary prior distribution (\ref{eqn:prior}) and the loss function (\ref{eqn:loss}), then the Bayes estimator for $\theta$ is 
    \begin{equation}\label{eqn:bayes}
    \hat{\theta}_{B} = x + \frac{ks^2 f_x(x,s^2) }{(k-2)f(x,s^2) - 2s^2 f_{s^2}(x,s^2)}.
    \end{equation}
\end{theorem}

\medskip
In equation~(\ref{eqn:bayes}), the Bayes estimator depends solely on the marginal distribution of the data and its first-order partial derivatives. 
When the variance \( \sigma^2 \) is assumed to be known—equivalently, when \( s^2 = \sigma^2 \) is a constant and \( k = \infty \)—the general form of the Bayes estimator in~(\ref{eqn:bayes}) simplifies to
\[
    \hat{\theta}_{B} 
    = x + \sigma^2 \frac{f_x(x, \sigma^2)}{f(x, \sigma^2)}.
\]
This expression is known as \emph{Tweedie’s formula}. The extension from the classical Tweedie framework---where the variances are assumed to be known and equal---to the present setting with heterogeneous and unknown variances. 
In particular, our formulation extends the classical empirical Bayes paradigm to encompass models in which both the means and variances are treated as random, thereby accommodating a broader range of practical scenarios. Notably, we impose no structural assumptions on the joint prior distribution of $(\theta, \sigma^2)$, allowing for arbitrary dependence between the mean and variance components. 

The Bayes estimator in equation~(\ref{eqn:bayes}) depends on the marginal probability density function of the data, which is typically unknown and therefore not directly available in practice. 
Nevertheless, it provides an idealized benchmark that achieves the minimum risk under the assumed model. 
To assess the performance of practical estimators, it is natural to compare them against this theoretical gold standard. 
In the following, we construct an empirical Bayes counterpart that seeks to mimic the Bayes estimator by replacing the unknown marginal quantities with their data-driven estimates.

\subsection{Estimation of the unknown quantities}\label{sec:esti:unknown}

To implement the proposed nonparametric $f$-modeling, we estimate the joint marginal density $f(x,s^2)$ and its first-order partial derivatives with respect to $x$ and $s^2$, which are required to evaluate the Bayes estimator in Theorem~3.1.
Since $s_i^2$ is bounded away from zero, we apply a logarithmic transformation and estimate the joint density of $(x_i,\log s_i^2)$ using a bivariate kernel density estimator with Gaussian kernel, implemented via the \texttt{ks} package in \textsf{R}.
For bandwidth selection, we compute marginal bandwidths $h_x$ and $h_s$ for $x_i$ and $\log s_i^2$, respectively, using Silverman's rule-of-thumb \citep[e.g.][]{Silverman:1986}, and construct the bandwidth matrix as $\mathrm{diag}(h_x,h_s)$.
This diagonal structure allows different smoothing scales along each coordinate while maintaining numerical stability.

The estimated density on the original scale $(x,s^2)$ is then obtained by $\widehat f(x,s^2)=\widehat f(x,\log s^2)/s^2$.
To compute the required partial derivatives, we employ symmetric finite-difference approximations.
Specifically, the derivative with respect to $x$ is approximated by $\widehat f_x(x,s^2)\approx\{\widehat f(x+e,s^2)-\widehat f(x-e,s^2)\}/2e$, and the derivative with respect to $s^2$ is approximated by $\widehat f_{s^2}(x,s^2)\approx\{\widehat f(x,s^2+e)-\widehat f(x,s^2-e)\}/2e$, for some $e<s^2$.
In our implementation, we set $e=\min(10^{-3}, s_i^2/2)$.

\subsection{Regret bound}
To quantitatively evaluate the discrepancy between an arbitrary estimator and the Bayes benchmark, we study \emph{regret} which measures the excess risk incurred by using a given estimator instead of the Bayes estimator and thus provides a natural criterion for assessing the efficiency of empirical Bayes procedures \citep{zhang1997empirical, Jiang:Zhang:2009}.

Formally, let \( \hat{\theta} \) be any estimator of \( \theta \) under a given loss function \( L(\cdot, \cdot) \). 
The regret is defined as
\[
\mathrm{Regret}(\hat{\theta}) 
= \mathbb{E} \left[ L(\hat{\theta}, \theta) - L(\hat{\theta}_B, \theta) \right],
\]
where \( \hat{\theta}_B \) denotes the Bayes estimator. 
A smaller regret indicates that the estimator \( \hat{\theta} \) performs closer to the Bayes optimal solution. 
In the next result, we establish a theoretical bound for this regret, which provides insight into the asymptotic optimality of the proposed empirical Bayes estimator.

\begin{lemma}\label{lemma:regret}
Let $\hat{\theta}$ be any estimator of $\theta$. Assume the loss function (\ref{eqn:loss}), then
\[
{\rm Regret}(\hat{\theta}) = \mathbb{E}\left( \frac{1}{\sigma^2}(\hat{\theta}-\hat{\theta}_B)^2\right).
\]
\end{lemma}

Next, we construct the empirical Bayes estimator of \( \theta \). 
Recall that the Bayes estimator involves the marginal density function \( f(x, s^2) \) and its partial derivatives \( f_x(x, s^2) \) and \( f_{s^2}(x, s^2) \). 
In practice, these quantities are unknown and are estimated from the observed data using methods such as the one provided in Section \ref{sec:esti:unknown}. 
Let $\widehat{f}(x, s^2)$, $\widehat{f}_x(x, s^2)$, and $\widehat{f}_{s^2}(x, s^2)$ denote consistent estimators of \( f(x, s^2) \), \( f_x(x, s^2) \), and \( f_{s^2}(x, s^2) \), respectively. 
Substituting these estimates into the Bayes estimator yields the empirical Bayes estimator
\[
\hat{\theta}_{\rm EB} 
= x + 
\frac{k s^2 \cdot \widehat{f}_x(x, s^2)}
{(k - 2)\widehat{f}(x, s^2) - 2 s^2 \cdot \widehat{f}_{s^2}(x, s^2)}.
\]

It is worth noting that the denominator in the above expression may approach zero or the numerator may become extremely small, particularly near the boundary of the support of \( (x, s^2) \), which can lead to numerical instability. 
To alleviate this issue, we introduce a tempered version of the estimator by imposing a lower bound \( \rho > 0 \) on the denominator. 
The resulting \emph{tempered empirical Bayes estimator} is defined as
\begin{equation}\label{eqn:embayes}
\hat{\theta}_{\rm EBT}(\rho)
= x +
\frac{k s^2 \cdot \widehat{f}_x(x, s^2)}
{\left\{ (k - 2)\widehat{f}(x, s^2) - 2 s^2 \cdot \widehat{f}_{s^2}(x, s^2) \right\} \vee \rho},
\end{equation}
where \( a \vee b = \max(a, b) \) ensures the denominator remains bounded away from zero.

\begin{theorem}\label{thm:regret}
Define $D_k(x, s^2)\equiv (k-2)f(x,s^2)-2s^2f_{s^2}(x,s^2)$ and $\widehat{D}_k(x, s^2)$ as the estimated counterpart. 
Assume that $\{D_k(x,s^2) \vee\rho\}^{-1}ks^2f_x(x,s^2)\le A(\rho)$. Then the regret of the tempered empirical Bayes estimator is bounded as
\[
\left\{ {\rm Regret}(\hat{\theta}_{\rm EBT}(\rho))\right\}^{1/2}\le \Delta_1+ \frac{A(\rho)}{\rho\sigma^2}\Delta_2 + \Delta_3, 
\]
where
\[
\Delta_1=  \frac{1}{\rho}\left\lVert \frac{s^2}{\sigma^2}\left\{\hat{f}_x(x,s^2)-f_x(x,s^2)\right\}\right\rVert_{f(x,s^2)},
  \ \ \ \ 
\Delta_2= \left\lVert \left\{\widehat{D}_k(x,s^2)\vee\rho\right\} - \left\{D_k(x,s^2)\vee\rho\right\} 
\right\rVert_{f(x,s^2)},
\]
and
$$
\Delta_3=\left( \iint \frac{k^2s^4}{\sigma^4}\left\{ \frac{f_x(x,s^2)}{D_k(x,s^2)}\right\}^2 \cdot f(x,s^2) \cdot \left\{1-\frac{D_k(x,s^2)}{\rho}\right\}_+^2 dxds^2 \right)^{1/2}.
$$
\end{theorem}

Theorem \ref{thm:regret} characterizes how estimation error in the marginal density propagates to excess risk. While sharper rates may be attainable under additional smoothness assumptions, the present bound suffices to establish asymptotic efficiency relative to the Bayes benchmark.

The proposed tempered empirical Bayes estimator is closely related to the regularized empirical Bayes framework developed by \cite{zhang1997empirical}. 
\cite{zhang1997empirical} proposed the idea of stabilizing the empirical Bayes estimator by truncating the denominator with a lower bound \( \rho>0 \), replacing \( \hat f \) by \( \hat f \vee \rho \) to avoid numerical instability when the estimated marginal density is small.  \cite{muralidharan2012high} extended the result to exponential families. 
Our Theorem~\ref{thm:regret} extends these ideas to the heteroscedastic setting by incorporating both \( x \) and \( s^2 \) into the marginal density.

%% file: posterior.tex
We now describe a practical procedure for approximating posterior distribution functions based on the MGF representation. The approach relies on discretization and entropy regularization, yielding a numerically stable approximation that performs well in practice. 


%
\subsection{Posterior cumulative distribution function}\label{sec:pos}
The posterior cumulative distribution function (CDF) of \(\theta_i\) can be expressed as
$$
F_{\theta_i}(z) \equiv \mathbb{P}(\theta_i \leq z \mid \mathcal{D})
= \mathbb{P}(U_i/V_i \leq z \mid \mathcal{D})
= \mathbb{P}(U_i - zV_i \leq 0 \mid \mathcal{D}),
$$
where $U_i$ and $V_i$ are defined in Section~\ref{sec:model}. Let 
$W_i(z) = U_i - z V_i|\mathcal{D}$.
The moment generating function of $W_i(z)$ is $M_{W_i(z)}(t) = M_{U_i,V_i}(z, -zt)$, which can be estimated from the marginal distribution of $(x_i, s_i^2)$.
This, in turn, yields an estimator of $F_{W_i(z)}(0) = \mathbb{P}\{W_i(z) \le 0\}$, i.e., the posterior probability $F_{\theta_i}(z)$.

Specifically, we approximate the distribution of $W_i(z)$ by a discrete measure supported on a finite set of grid points. Let \(a_1, \ldots, a_S\) be a collection of knots, and approximate the distribution of $W_i(z)$ by
\[
\sum_{s=1}^S p_s \, \delta_{a_s},
\]
where \(\delta_{a_s}\) denotes the Dirac measure at \(a_s\). 
A natural choice of the grid points \(\{a_1, \ldots, a_S\}\) is a uniform partition over the interval 
\[
\Big[
\mathbb{E}(W_i(z)) - c_w\sqrt{{\rm Var}(W_i(z))},
\ \mathbb{E}(W_i(z)) + c_w\sqrt{{\rm Var}(W_i(z))}
\Big],
\]
where we set $c_w=5$ in our implementation. 
The above interval covers a wide and representative range of the posterior mass.

Given multiple values of the MGF \(M_{W_i(z)}(t)\) evaluated at \(t=t_1, \ldots, t_L\), 
the corresponding discrete probabilities \(\{p_s\}\) can be determined by solving the following constrained optimization problem:
\[
\min_{p_1, \ldots, p_S} \sum_{s=1}^S p_s \log p_s, 
\quad \text{subject to} \quad 
\sum_{s=1}^S p_s \exp(t_l a_s) = M_{W_i(z)}(t_l), \ \ l = 1, \ldots, L.
\]
The solution takes the exponential family form
\begin{equation}\label{eq:discrete-pos}
p_s(\lambda)
= \frac{\exp \big(\lambda^\top H_s\big)}
{\sum_{s'=1}^S \exp \big(\lambda^\top H_{s'}\big)},
\end{equation}
where \(H_s = \big(\exp(t_1 a_s) - M_{W_i(z)}(t_1), \ldots, \exp(t_L a_s) - M_{W_i(z)}(t_L)\big)\),
and \(\lambda = (\lambda_1, \ldots, \lambda_L)\) denotes the vector of Lagrange multipliers 
enforcing the moment-matching constraints
\[
\sum_{s=1}^S H_s \exp(\lambda^\top H_s) = 0.
\]
The Lagrange multipliers can be efficiently estimated via the Newton–Raphson algorithm, 
which updates the current iterate \(\lambda_{(t)}\) as
\[
\lambda_{(t+1)} 
\leftarrow 
\lambda_{(t)} -
\Bigg\{
\sum_{s=1}^S H_s H_s^\top \exp\big(\lambda_{(t)}^\top H_s\big)
\Bigg\}^{-1}
\sum_{s=1}^S H_s \exp\big(\lambda_{(t)}^\top H_s\big).
\]
Let \(\tilde{\lambda}\) denote the converged solution. 
Substituting \(\lambda = \tilde{\lambda}\) into (\ref{eq:discrete-pos}) yields 
the discrete approximation of the distribution of \(W_i(z)\).
Then, the posterior probability can be evaluated as
\begin{equation}\label{eqn:post:prob}
F_{W_i(z)}(0)  \approx
\sum_{s=1}^S p_s I(a_s \leq 0).
\end{equation}

In the traditional $f$-modeling approach, traditional approaches focus on the estimation of the posterior moments without estimating the prior distribution. 
In the $g$-modeling framework, the prior distribution is estimated directly, while its posterior distribution remains unclear.
In contrast, our framework recovers the full posterior distribution of \(\theta_i\) based on the $f$-modeling. 
By an $f$-modeling representation of the MGF, 
our method enables comprehensive inference including 
confidence intervals and multiple testing, as described in the subsequent section.

\begin{remark}
    The proposed procedure can be viewed as an inverse problem that reconstructs distributional functionals from estimated moment generating functions under discretization and entropy regularization. The behavior of the estimator is governed by the quality of MGF approximation on a bounded domain and by the regularizing role of the maximum-entropy formulation, which stabilizes the inversion in practice. Theoretical developments in this direction are an interesting topic for future research. In the present work, we complement these insights with extensive simulations and real-data analyses, which demonstrate that the method is numerically stable and achieves substantial improvements over existing empirical Bayes approaches across a wide range of settings.
\end{remark}

\begin{remark}
The proposed method can be extended to much broader settings. Let $g(\theta, \sigma^2)$ be any parameter of interest. Then the corresponding posterior quantity can be written as
\[
g(\theta, \sigma^2 \mid \mathcal{D}) = g(U/V,\, 1/V).
\]
Similarly, a bivariate discretization scheme can be used to estimate the joint distribution of $(U, V)$ and thereby conduct inference for $g(\theta, \sigma^2)$. One can easily obtain an estimator of $\mathbb{E}g(\theta,\sigma^2|\mathcal{D})$ based on this framework. As an application, it provides another way to estimate $\mathbb{E}\frac{U}{V}|\mathcal{D}$ and $\mathbb{E}\frac{1}{V}|\mathcal{D}$.

In the current section, our goal is to estimate the posterior probability 
\[
\mathbb{P}(\theta \le z \mid \mathcal{D})
= \mathbb{E}\!\left[ I(U - V z \le 0) \right],
\]
where $I(\cdot)$ denotes the indicator function. 
By defining the transformed variable $W = U - Vz$, this bivariate problem is reduced to a univariate one, enabling more efficient computation and analysis.

\end{remark}

\subsection{Empirical Bayes confidence intervals}

Using the estimated posterior distribution function $F_{\theta_i}(z)$ given in Section~\ref{sec:pos}, we construct empirical Bayes confidence intervals for $\theta_i$ at nominal level $1 - \alpha$, defined as the quantile interval $(z_L,z_U)$ satisfying $F_{\theta_i}(z_L)=\alpha/2$ and $F_{\theta_i}(z_U)=1-\alpha/2$.
These quantiles are computed numerically via a bisection search over $z$, exploiting the monotonicity of the cumulative distribution function $F_{\theta_i}(z)$.

Unlike classical confidence intervals based solely on the sampling distribution of $x_i$, the proposed interval construction directly reflects the estimated posterior uncertainty of $\theta_i$, which incorporates information borrowing across observations through the joint marginal distribution $f(x,s^2)$.
As a result, the resulting intervals naturally adapt to heterogeneity in both signal strength and variance, leading to shorter intervals while maintaining appropriate uncertainty quantification.
It is worth emphasizing that, although the posterior distribution is approximated through discretization, the quantile-based confidence intervals depend only on the estimated distribution function $F_{\theta_i}(z)$, rather than on any parametric assumptions on the prior, as assumed in \cite{Hwang:Qiu:Zhao:2009}.
Consequently, the proposed empirical Bayes confidence intervals remain valid under a broad class of joint prior distributions 
and provide a coherent uncertainty quantification framework that extends beyond point estimation.

\subsection{Multiple Hypothesis Testing}\label{sec:testing}

It is natural to extend the proposed framework to hypothesis testing problems, where the goal is to make decisions about whether each parameter \(\theta_i\) belongs to a specified null region. Consider the general form of hypotheses:
\begin{equation}\label{eqn:hypothesis}
    H_i^0: \theta_i \in \Theta_0, 
    \qquad 
    H_i^1: \theta_i \in \Theta_0^c,
\end{equation}
where \(\Theta_0\subset \mathbb{R}\) denotes the null parameter space.  
This formulation encompasses a wide range of testing problems as special cases that have been studied in the empirical Bayes and decision-theoretic literature.
For example, \cite{ruppert2007exploring} considered 
\(\Theta_0 = \{\theta_i: |\theta_i| \le \delta\}\),
which defines an indifference region around zero, representing parameters that are too small to be considered practically significant.
This formulation has been further discussed and extended in 
\cite{morris2008comment, van2007estimating}, 
and more recently in \cite{xiang2024interpretation}.  
Another important example, proposed by \cite{gelman2000type} and \cite{stephens2017false}, 
takes \(\Theta_0 = (-\infty, 0)\),  
which corresponds to the problem of sign recovery—determining whether the effect size is positive or negative.

Under a weighted 0–1 loss function, the Bayes-optimal decision rule is determined by the posterior probability that the parameter falls in the null region,
\[
PN_i = \mathbb{P}(\theta_i \in \Theta_0 \mid \mathcal{D}),
\]
which we refer to as the posterior null probability.  
Intuitively, small values of \(PN_i\) provide strong evidence against the null, while large values suggest insufficient evidence to reject it.
When \(\Theta_0 = \{\theta_i: |\theta_i| \le \delta\}\),  
the posterior null probability is
\[
PN_i
= \mathbb{P} \left(-\delta \le \frac{U_i}{V_i} \le \delta \,\middle|\, \mathcal{D}\right)
= \mathbb{P}(U_i - \delta V_i \le 0 \mid \mathcal{D})
  - \mathbb{P}(U_i + \delta V_i \le 0 \mid \mathcal{D}).
\]
Both quantities on the right-hand side involve posterior probabilities of the form
\(\mathbb{P}(U_i - z V_i \le 0 \mid \mathcal{D})\),
which can be estimated accurately using the MGF-based discretization method developed in Section~\ref{sec:pos}.  
Hence, the proposed empirical Bayes approach provides a direct and unified mechanism for computing posterior probabilities and thereby constructing optimal testing rules under general null regions \(\Theta_0\).

The key innovation here is that, unlike traditional Tweedie-type empirical Bayes methods—which yield only the posterior mean of \(\theta_i\) and hence cannot directly quantify uncertainty or support hypothesis testing—the proposed MGF-based framework reconstructs the full posterior distribution of each \(\theta_i\).  
As a result, it enables the direct computation of posterior tail probabilities and confidence regions, leading to data-driven Bayesian tests with valid FDR control under very general prior assumptions \citep{Efron:2004, Sun:Cai:2007}.

%% file: simulation.tex
\subsection{Empirical Bayes estimation}\label{sec:sim-EB}

Here we evaluate the performance of the point estimation through Monte Carlo simulation studies. 
Given the true value of $\theta_i$ and $\sigma_i^2$, the observed value $X_i$ and $S_i$ are generated according to the model (\ref{eqn:model}).
We considered the following data generating scenarios for $\theta_i$ and $\sigma_i^2$: 
\begin{align*}
{\rm (S1)}\ \ \ 
&\theta_i|z_i\sim z_i \mathcal{N}(-1,0.7^2) + (1-z_i)\mathcal{N}(-1+\eta/2,0.7^2), \ \ \ \ \sigma_i^2=\max(0.2, \sigma_{i\ast}^2), \\
& \sigma_{i\ast}^2|z_i\sim z_i{\rm Ga}(1,1) + (1-z_i)
{\rm Ga}(1+\eta/3,1), \ \ \ \mathbb{P}(z_i=1)=1-\mathbb{P}(z_i=0)=0.3,\\
{\rm (S2)}\ \ \ 
&\theta_i|z_i\sim z_i\eta_0 + (1-z_i) U(\eta/3, 2+\eta/3), \ \ \ \ \sigma_i^2\sim z_iU(0.5, 1/5)+(1-z_i)U(3,7), \\
&  \mathbb{P}(z_i=1)=1-\mathbb{P}(z_i=0)=0.4,  \\
{\rm (S3)}\ \ \ &\theta_i\sim 0.5\mathcal{N}(0, (0.5)^2) + 0.5 {\rm Ga}(2+\eta/3, 2), \ \ \ \ 
\log(\sigma_i^2)|\theta_i \sim \mathcal{N}(\theta_i/2, (0.2)^2),
\end{align*}
where $\eta$ is a shift parameter controlling the complexity of the data generating distribution.
Overall, the data generating process can be well approximated by parametric models when $\eta$ is close to $0$. 
In particular, when $\eta=0$ in (S1), the data generating distribution for $\theta_i$ and $\sigma_i^2$ are normal and gamma distributions, respectively.  
For each scenario, we adopted four combinations of $n\in \{500, 1000\}$ (sample size) and $k\in \{10, 20\}$ (degrees of freedom).

\begin{figure}[htb!]
\centering
\includegraphics[width=0.32\linewidth]{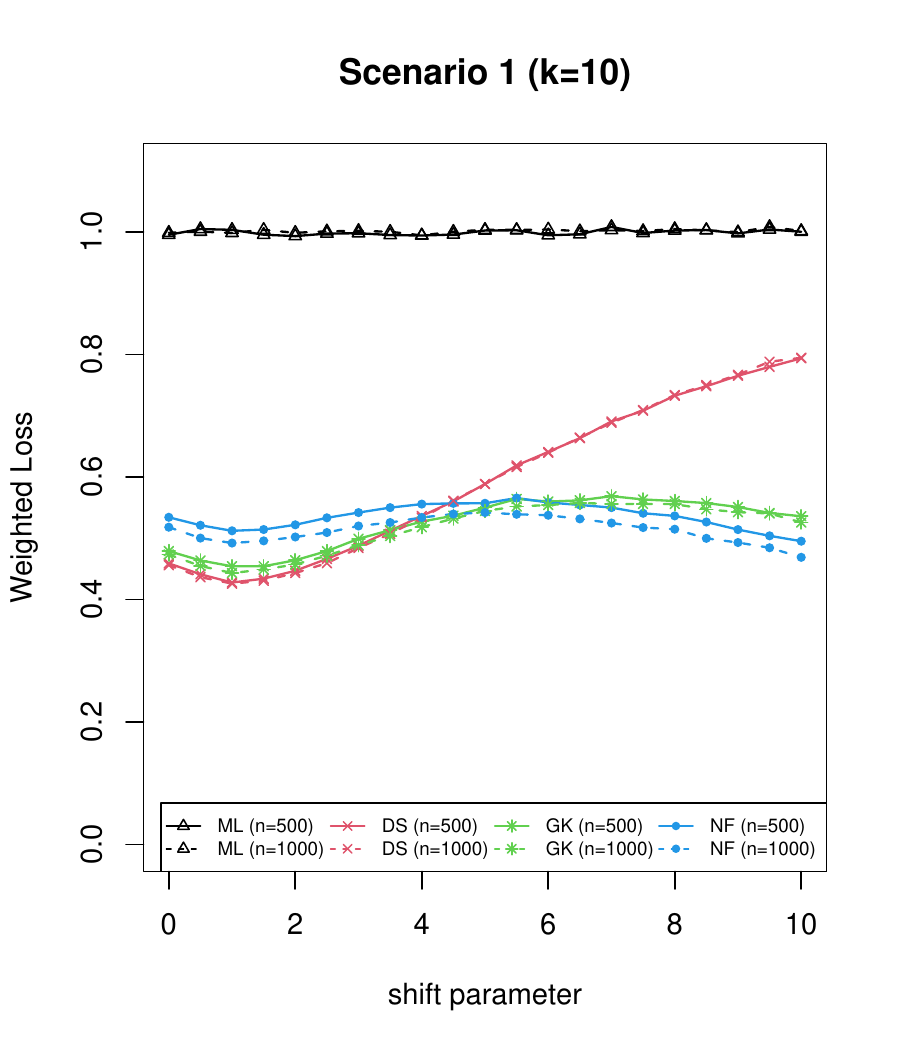}
\includegraphics[width=0.32\linewidth]{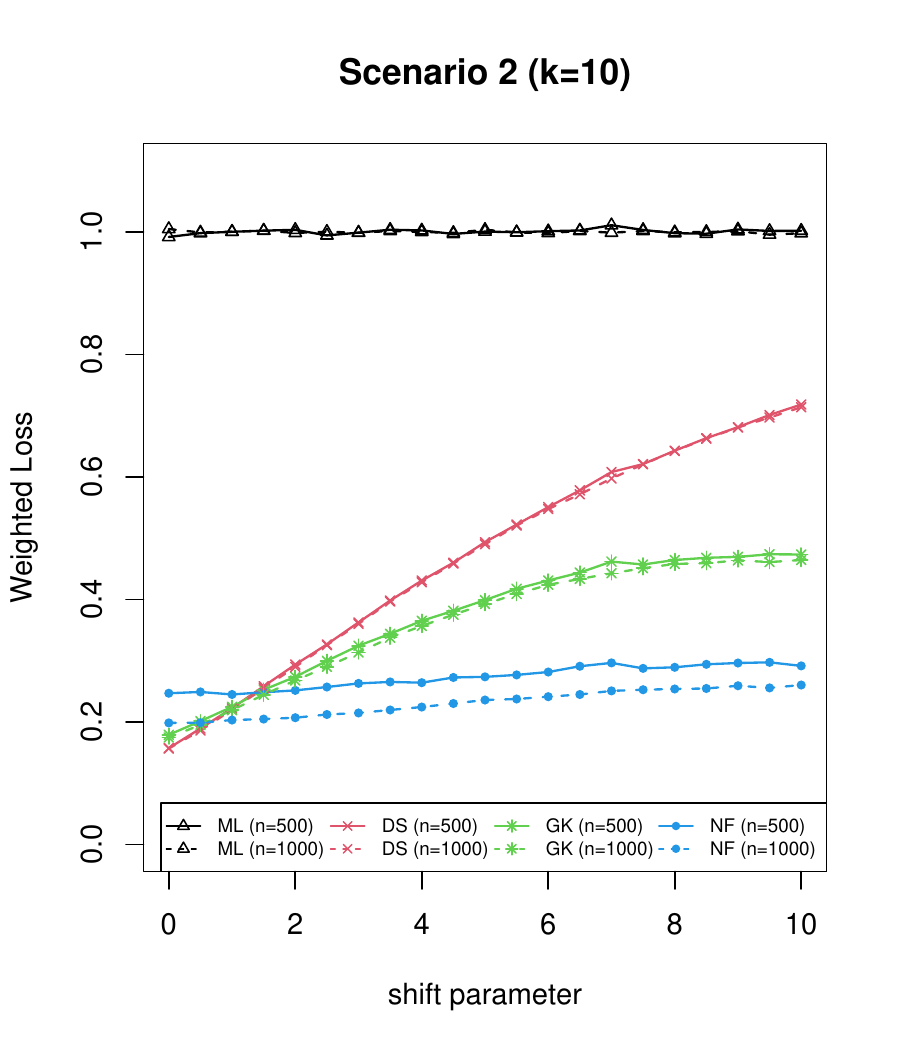}
\includegraphics[width=0.32\linewidth]{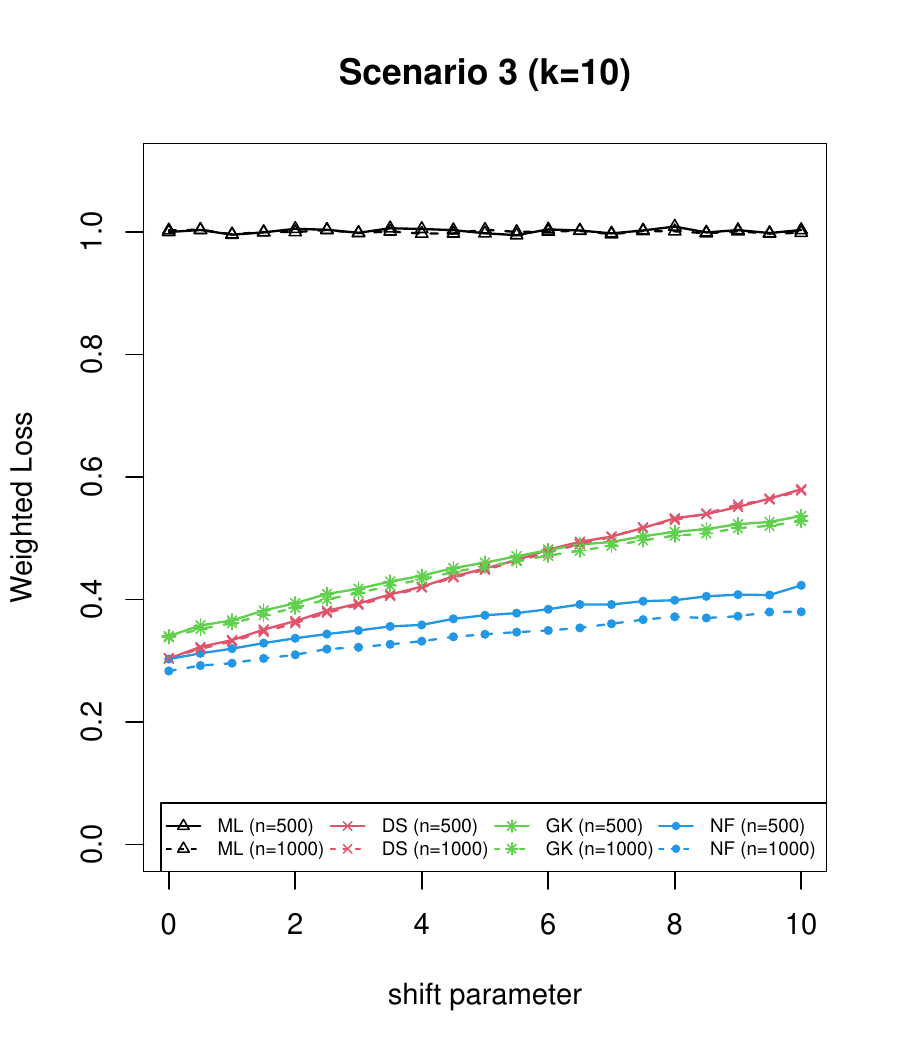}\\
\includegraphics[width=0.32\linewidth]{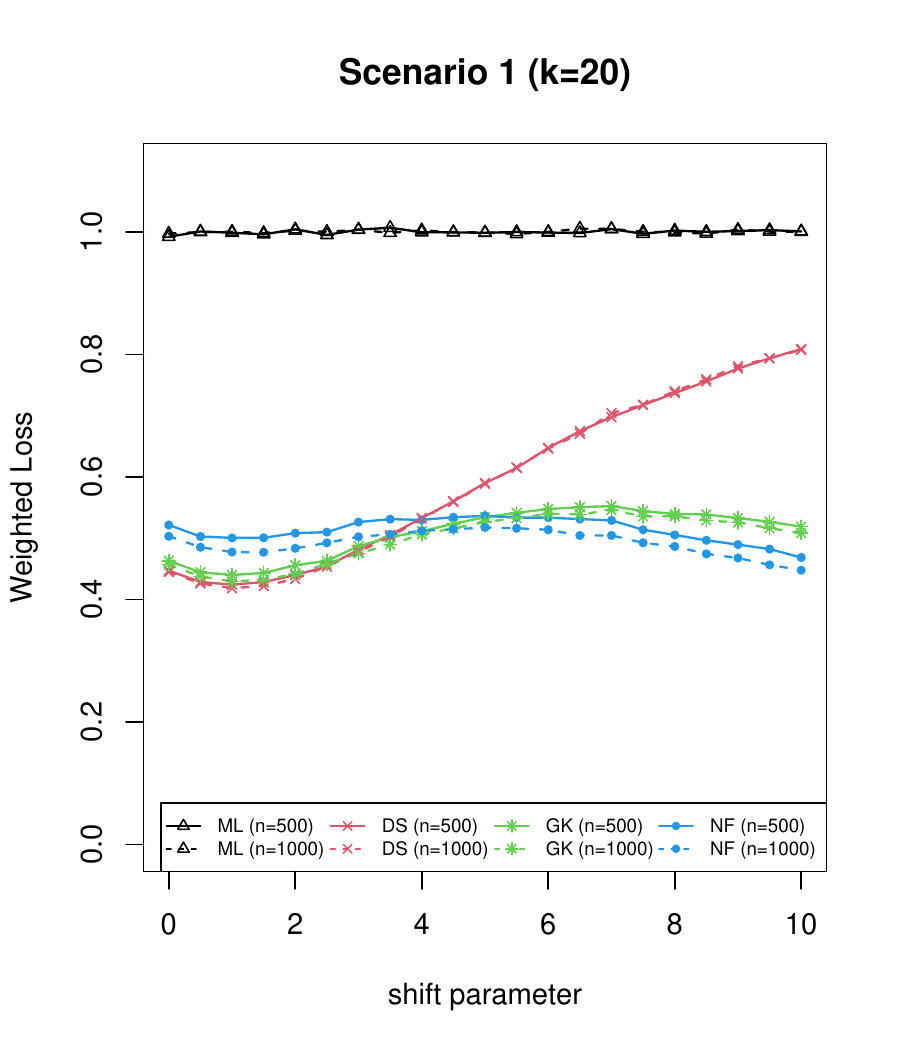}
\includegraphics[width=0.32\linewidth]{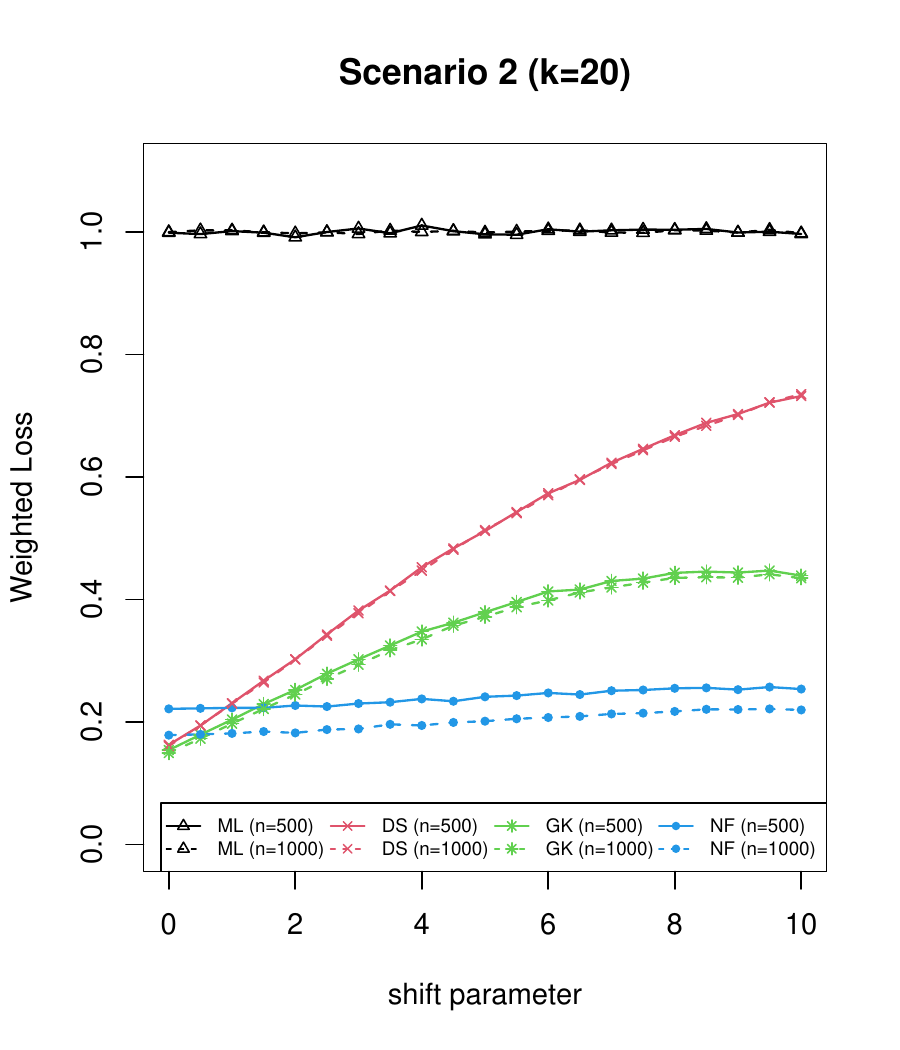}
\includegraphics[width=0.32\linewidth]{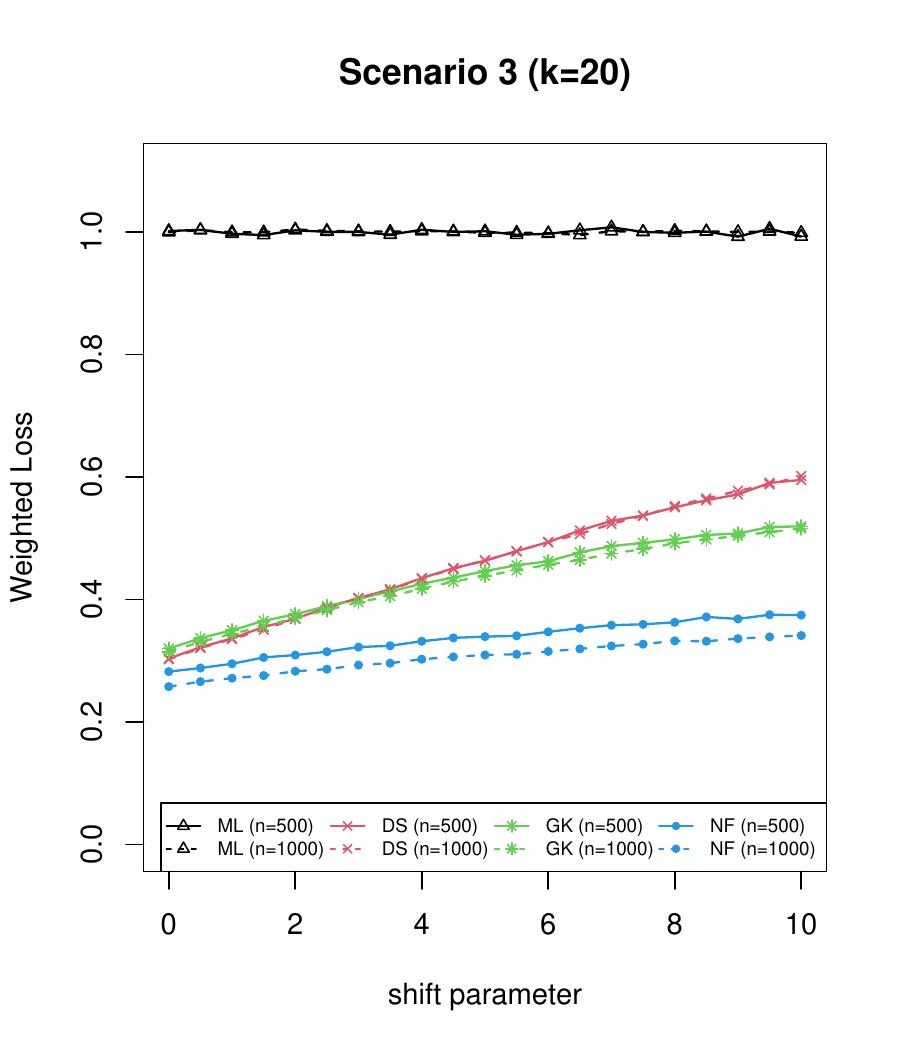}
\caption{ Weighted loss of ML, DS, GK and NF (proposal) under three data generating processes, averaged over 500 Monte Carlo replications. } 
\label{fig:wMSE}
\end{figure}

For the generated dataset, we applied the naive maximum likelihood (ML) estimator that uses the observed data $x_i$ directly, the parametric double-shrinkage (DS) estimator assuming a parametric hierarchical model, as proposed by \cite{Zhao:2010}, the heteroscedastic empirical Bayes estimator by \cite{gu2017empirical}, denoted by GK, and the proposed nonparametric $f$-modeling empirical Bayes estimator, denoted by NF. 
For the implementation of GK, we employed ``REBayes" package \citep{gu2017empirical}.
In NF, we used the EBT estimator by setting $\rho=10^{-3}$. 
We evaluated the estimation accuracy via the weighted loss, defined as $n^{-1}\sum_{i=1}^n \sigma_i^{-2}(\hat{\theta}_i-\theta_i)^2$.
The results under $\eta\in [0,10]$ are shown in Figure~\ref{fig:wMSE}, where the reported values are averaged over 300 Monte Carlo replications.

Since the theoretical loss of ML is equal to 1, and the empirical values closely match this benchmark.
Overall, the proposed NF consistently achieves the smallest loss across scenarios.
When $\eta$ is small, the data generating process is relatively simple and close to a parametric model; accordingly, all shrinkage estimators (DS, GK, and NF) show comparable performance.
However, as $\eta$ increases and the data distribution becomes more complex, the advantage of NF becomes pronounced, with significantly smaller loss than the competing methods.
In scenario (S3), the difference among the methods becomes more informative.
The GK method assumes a flexible model for the marginal distribution of the means and thus performs well when the heterogeneity arises mainly from variations in the distribution of $\theta_i$.
However, GK does not model the dependence between $\theta_i$ and $\sigma_i^2$, which is strongly present in this scenario.
In contrast, the parametric DS method imposes a simple parametric form on the distribution of $\theta_i$, but does account for the correlation between $\theta_i$ and $\sigma_i^2$.
As a result, DS achieves performance comparable to GK under (S3), whereas the proposed NF method continues to exhibit stable and superior performance.
This highlights the advantage of our two-dimensional $f$-modeling approach, which flexibly captures the joint structure of $(x_i, s_i^2)$ without restrictive modeling assumptions.

\subsection{Empirical Bayes confidence intervals}

We next evaluate the performance of the empirical Bayes confidence intervals. 
In this study, we employed the same data generating process and set $\eta=4$ in the three scenarios given in the previous section, and the same settings of $n$ and $k$. 
Based on the discretization approach described in Section~\ref{sec:interval}, we construct $95\%$ confidence intervals of $\theta_i$ by setting $S=100$ (the number of grid for discrete approximation) and $(t_1,\ldots,t_6)=(-0.3, -0.2, -0.1, 0.1, 0.2, 0.3)$ (the evaluation points of MGF).
For comparison, we applied the $t$-interval with the maximum likelihood estimator (observations), $(x_i, s_i^2)$, defined as $(x_i-t_k(0.025)s_i, x_i+t_k(0.025)s_i)$, where $t_k(0.025)$ denotes the upper $2.5\%$ quantile of the $t$-distribution with $k$ degrees of freedom.  
This interval is denoted by ML. 
Moreover, we adopt a parametric double shrinkage (DS) confidence interval with unequal and unknown variances, as proposed by \cite{Hwang:Qiu:Zhao:2009}.

Based on $M=50$ Monte Carlo replications, we evaluated overall coverage probability (CP) and average length (AL), defined as 
$$
{\rm CP}=\frac{1}{Mn}\sum_{m=1}^M\sum_{i=1}^n I\Big(\theta_i^{(m)}\in {\rm CI}_i^{(m)}\Big), \ \ \ \ \ 
{\rm AL}=\frac{1}{Mn}\sum_{m=1}^M\sum_{i=1}^n \big|{\rm CI}_i^{(m)}\big|,
$$
where $\theta_i^{(m)}$ denotes the true value and ${\rm CI}_i^{(m)}$ is the confidence interval in the $m$th replication.
The resulting values are reported in Table~\ref{tab:sim-CI}.
It shows that all three methods attain CP values close to the nominal level across all scenarios. 
Compared with the naive ML intervals, both DS and NF produce substantially shorter confidence intervals, reflecting the efficiency gains from shrinkage. 
Comparing DS and NF, NF tends to yield more efficient intervals than DS when the degrees of freedom gets larger, achieving shorter average lengths while maintaining comparable CP values.

\begin{table}[t]
\centering
\caption{Coverage probability (CP) and average length (AL) of $95\%$ confidence intervals of $\theta_i$ under $n=500$, averaged over 50 Monte Carlo replications. }
\label{tab:sim-CI}

\medskip
\begin{tabular}{ccccccccccccccc}
\hline
&&&& \multicolumn{3}{c}{CP (\%)} && \multicolumn{3}{c}{AL}\\
$n$ & $k$ & Scenario  &  & ML & DS & NF &  & ML & DS & NF \\
\hline
 &  & 1 &  & 95.1 & 96.6 & 96.2 &  & 6.55 & 5.12 & 5.57 \\
500 & 10 & 2 &  & 94.8 & 96.9 & 98.5 &  & 7.51 & 5.58 & 6.06 \\
 &  & 3 &  & 94.9 & 95.9 & 95.0 &  & 6.07 & 4.43 & 4.71 \\
\hline
&  & 1 &  & 94.9 & 96.1 & 95.7 &  & 6.18 & 4.99 & 4.68 \\
500 & 20 & 2 &  & 95.1 & 96.5 & 98.5 &  & 7.11 & 5.45 & 4.98 \\
 &  & 3 &  & 94.8 & 95.5 & 95.9 &  & 5.78 & 4.29 & 4.11 \\
\hline
&  & 1 &  & 95.1 & 96.8 & 96.5 &  & 6.56 & 5.11 & 5.53 \\
1000 & 10 & 2 &  & 95.1 & 97.0 & 98.9 &  & 7.51 & 5.60 & 5.98 \\
 &  & 3 &  & 95.1 & 96.2 & 95.7 &  & 6.09 & 4.45 & 4.70 \\
\hline
&  & 1 &  & 95.2 & 96.5 & 96.2 &  & 6.20 & 4.98 & 4.66 \\
1000 & 20 & 2 &  & 95.0 & 96.7 & 99.0 &  & 7.12 & 5.46 & 4.93 \\
 &  & 3 &  & 95.0 & 95.8 & 96.6 &  & 5.77 & 4.32 & 4.07 \\

\hline
\end{tabular}
\end{table}

\subsection{Multiple testing}

We next evaluate the performance of multiple testing under the proposed method compared with the naive approach using the observed data $(x_i, s_i^2)$.
We employ the same three scenarios used in Section~\ref{sec:sim-EB} with $\eta=4$ in this study. 
We set $\delta=1$ throughout this study, that is, $\theta_i$'s satisfying $|\theta_i|\leq  1$ are regarded as null signals.
Regarding the sample size and the degrees of freedom, we set $k=10$ and $n=1000$.
For the generated observations, we applied the null posterior probability based on kernel density estimation (NF-NP), described in Section~\ref{sec:testing}).
For comparison, we employ the frequentist testing procedure for the null hypothesis, $H_{0i}: |\theta_i|\leq \delta$.
Based on the observation, $(x_i, s_i^2)$, the $p$-value of $H_{0i}$ can be obtained as 
$$
p_i = 2\left\{1 - F_{t_k}\left(\frac{|x_i| - \delta}{s_i}\right)\right\},
$$
where $F_{t_k}(\cdot)$ denotes the cumulative distribution function of the $t$-distribution with $k$ degrees of freedom. 
This procedure tests the null hypothesis that the true effect lies within the interval $[-\delta, \delta]$ by evaluating whether the observed signal exceeds the predefined null region after accounting for sampling variability.
By using the $p$-values, we apply the same procedure for FDR control as used in NF.
This corresponds to directly using the observed data without any shrinkage, which will be denoted by ``T-test''.
On the other hand, instead of the raw sampling variance $s_i^2$, we first apply ``LIMMA'' package \citep{ritchie2015limma} to obtain shrinkage estimation of $\sigma_i^2$ and use it to approximate the null probability as $t_k(x_i, \hat{\sigma}_i)$, where $\hat{\sigma}_i$ is the shrinkage estimate produced by LIMMA package. 
Moreover, using the null probability of T-test, we apply the two-component mixture for the null probability, as proposed in \cite{strimmer2008unified}, and obtain the local FDR (shrinkage estimator of the null probability), denoted by LFDR. 
Finally, we simply applied the Benjamini-Hochberg procedure \citep{benjamini1995controlling} for the $p$-values, denoted by BH.
Under three FDR levels, $\alpha\in \{0.05, 0.1, 0.2\}$, the non-null signals are detected and empirical false discovery rate can be computed. 
In Table~\ref{tab:sim-FDR}, average false discovery rate based on 100 Monte Carlo replications are reported.

Overall, the proposed NF method performs quite well in all the scenarios since the empirical FDR is smaller than the nominal level and TPR and F1 values are tend to be larger than the other methods. 
Regarding the naive approaches, it can be seen that FDR tends to be too conservative or liberal depending on the scenarios.
For scenarios where FDRs of all the methods are controlled, TPR of the proposed NF method tends to be larger than the other methods, indicating the efficiency of the NF method.

\begin{table}[t]
\centering
\caption{False discovery rate (FDR) and F1 under $\alpha\in \{0.05, 0.1, 0.2\}$ of NF-NP (proposed), T-test, LIMMA, MIX and BH methods, averaged over 100 Monte Carlo replications. }
\label{tab:sim-FDR}

\medskip
{\small 
\begin{tabular}{ccccccccccccccc}
\hline
 &  &  &  & $\alpha=0.05$ &  &  &  & $\alpha=0.1$ &  &  &  & $\alpha=0.2$ &  \\
Scenario &  &  & FDR & TPR & F1 &  & FDR & TPR & F1 &  & FDR & TPR & F1 \\
\hline
 & NF-NP &  & 2.3 & 42.9 & 59.1 &  & 3.9 & 73.7 & 83.3 &  & 12.1 & 97.5 & 92.4 \\
 & T-Test &  & 3.3 & 39.0 & 55.6 &  & 5.2 & 56.5 & 70.7 &  & 8.9 & 78.9 & 84.6 \\
1 & LIMMA &  & 2.5 & 39.5 & 56.2 &  & 4.7 & 57.4 & 71.6 &  & 9.0 & 79.8 & 85.0 \\
 & MIX &  & 0.0 & 0.2 & 0.5 &  & 0.3 & 1.0 & 1.9 &  & 0.6 & 4.7 & 8.8 \\
 & BH &  & 0.3 & 0.9 & 1.8 &  & 0.6 & 4.2 & 7.9 &  & 1.3 & 14.9 & 25.7 \\
 \hline
 & NF-NP &  & 0.4 & 73.9 & 84.6 &  & 3.1 & 95.9 & 96.3 &  & 16.6 & 99.0 & 90.5 \\
 & T-Test &  & 5.2 & 41.8 & 58.0 &  & 9.8 & 60.3 & 72.3 &  & 20.1 & 79.8 & 79.8 \\
2 & LIMMA &  & 3.3 & 47.6 & 63.7 &  & 8.4 & 65.3 & 76.2 &  & 20.1 & 82.2 & 81.0 \\
 & MIX &  & 0.0 & 0.2 & 0.3 &  & 0.0 & 0.2 & 0.3 &  & 0.0 & 0.2 & 0.3 \\
 & BH &  & 0.0 & 0.3 & 0.5 &  & 0.5 & 1.3 & 2.5 &  & 1.0 & 7.7 & 14.1 \\
 \hline
 & NF-NP &  & 4.6 & 17.5 & 29.2 &  & 5.1 & 34.8 & 50.3 &  & 10.3 & 65.5 & 75.2 \\
 & T-Test &  & 16.9 & 25.4 & 38.8 &  & 23.2 & 41.0 & 53.4 &  & 33.2 & 62.1 & 64.3 \\
3 & LIMMA &  & 12.2 & 29.7 & 44.3 &  & 20.3 & 45.1 & 57.5 &  & 32.2 & 64.9 & 66.3 \\
 & MIX &  & 15.0 & 0.2 & 0.0 &  & 15.0 & 0.2 & 0.0 &  & 15.0 & 0.2 & 0.0 \\
 & BH &  & 0.0 & 0.0 & 0.1 &  & 0.8 & 0.1 & 0.2 &  & 1.6 & 0.5 & 0.9 \\
\hline
\end{tabular}
}
\end{table}

%% file: realdata.tex
We demonstrate the proposed method using the baseball batting data available in the \texttt{REBayes} package in \textsf{R} \citep{REBayes}.  
For each player $i$, the dataset contains aggregated batting records over multiple half-season periods, indexed by $j=1,\ldots,n_i$.  
For each half-season $j$, the numbers of hits and at-bats are denoted by $H_{ij}$ and $AB_{ij}$, respectively.
We apply the arcsine--square-root variance-stabilizing transformation
$$
x_{ij} = \arcsin \sqrt{\frac{H_{ij}+0.25}{AB_{ij}+0.5}},
$$
and combine the transformed observations $\{x_{ij}\}$ using inverse-variance weights $w_{ij}=4AB_{ij}$, which correspond to the inverse of the approximate variance under the binomial model.  
For each player $i$, we compute the weighted mean, weighted variance, and degrees of freedom as
$$
\bar{x}_i = \frac{\sum_{j=1}^{n_i} w_{ij} x_{ij}}{\sum_{j=1}^{n_i} w_{ij}}, \qquad
s_i^2 = \frac{\sum_{j=1}^{n_i} w_{ij}(x_{ij}-\bar{x}_i)^2}{(n_i-1)\sum_{j=1}^{n_i} w_{ij}}, \qquad
k_i = n_i - 1.
$$
These summary statistics are subsequently used as inputs for empirical Bayes estimation. 

We note that the effective degrees of freedom $k_i$ vary across players and ranges from 3 to 21.
As a result, a direct application of kernel density estimation to the pooled dataset $\{(x_i,s_i^2)\}_{i=1}^n$ may be inappropriate, since the marginal distributions are no longer identically distributed across observations.  
To address this heterogeneity, we estimate the conditional density $f(x_i,s_i^2\mid k_i)$ using the identity
$$
f(x_i,s_i^2\mid k_i)=\frac{f_J(x_i,s_i^2,k_i)}{f_M(k_i)},
$$
where $f_J(x_i,s_i^2,k_i)$ denotes the joint density of $(x_i,s_i^2,k_i)$ and $f_M(k_i)$ is the marginal density of $k_i$.  
Although $k_i$ take natural values, we treat it as a continuous variable in this application.  
Then, both densities are estimated via kernel density estimators with rule-of-thumb bandwidth selection.  
Once $f(x_i,s_i^2\mid k_i)$ is obtained, the Bayesian estimator defined in (\ref{eqn:bayes}) and the interval construction method described in Section~\ref{sec:interval} can be applied without further modification.

\begin{figure}[htb!]
\centering
\includegraphics[width=1.1\linewidth]{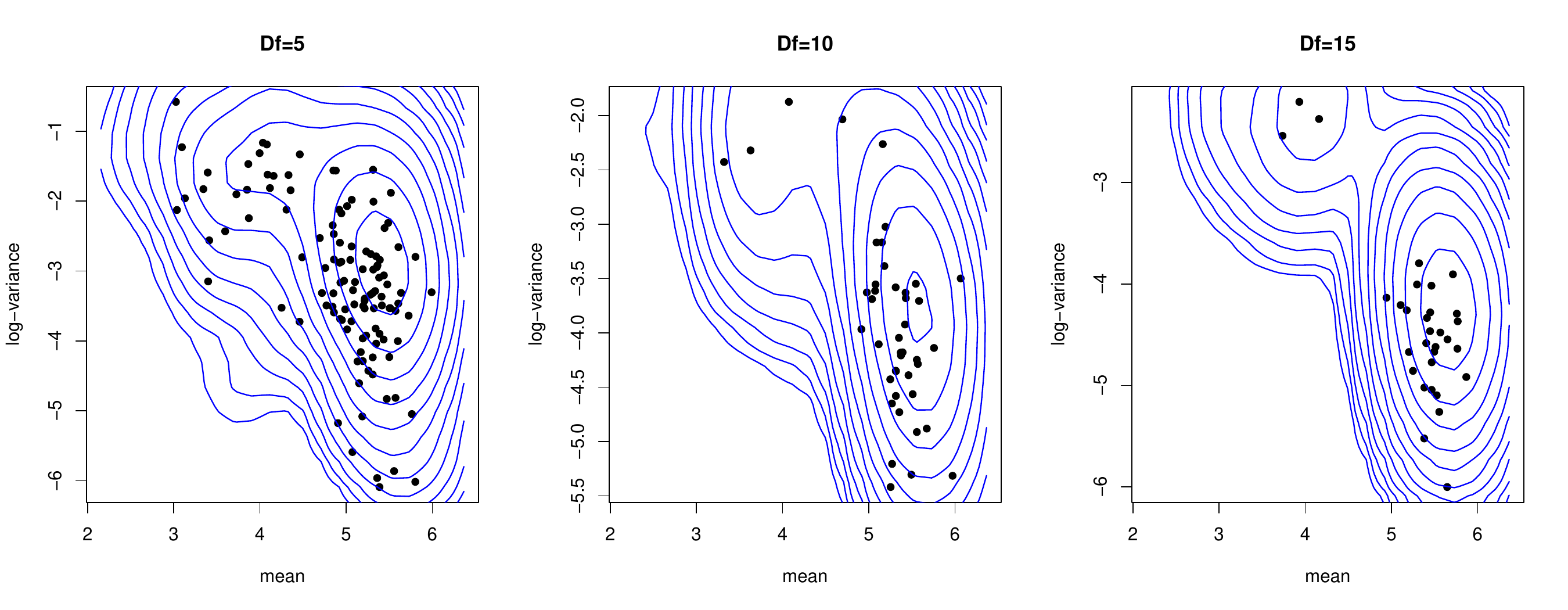}
\caption{ Estimated kernel density of $(x_i, \log s_i^2)$ conditional on the degrees of freedom $k_i=k\in \{5, 10, 15\}$. 
The dotted points are observations whose $k_i$ is included in the interval $[k-1, k+1]$.  } 
\label{fig:app-KDE}
\end{figure}

\begin{figure}[htb!]
\centering
\includegraphics[width=0.8\linewidth]{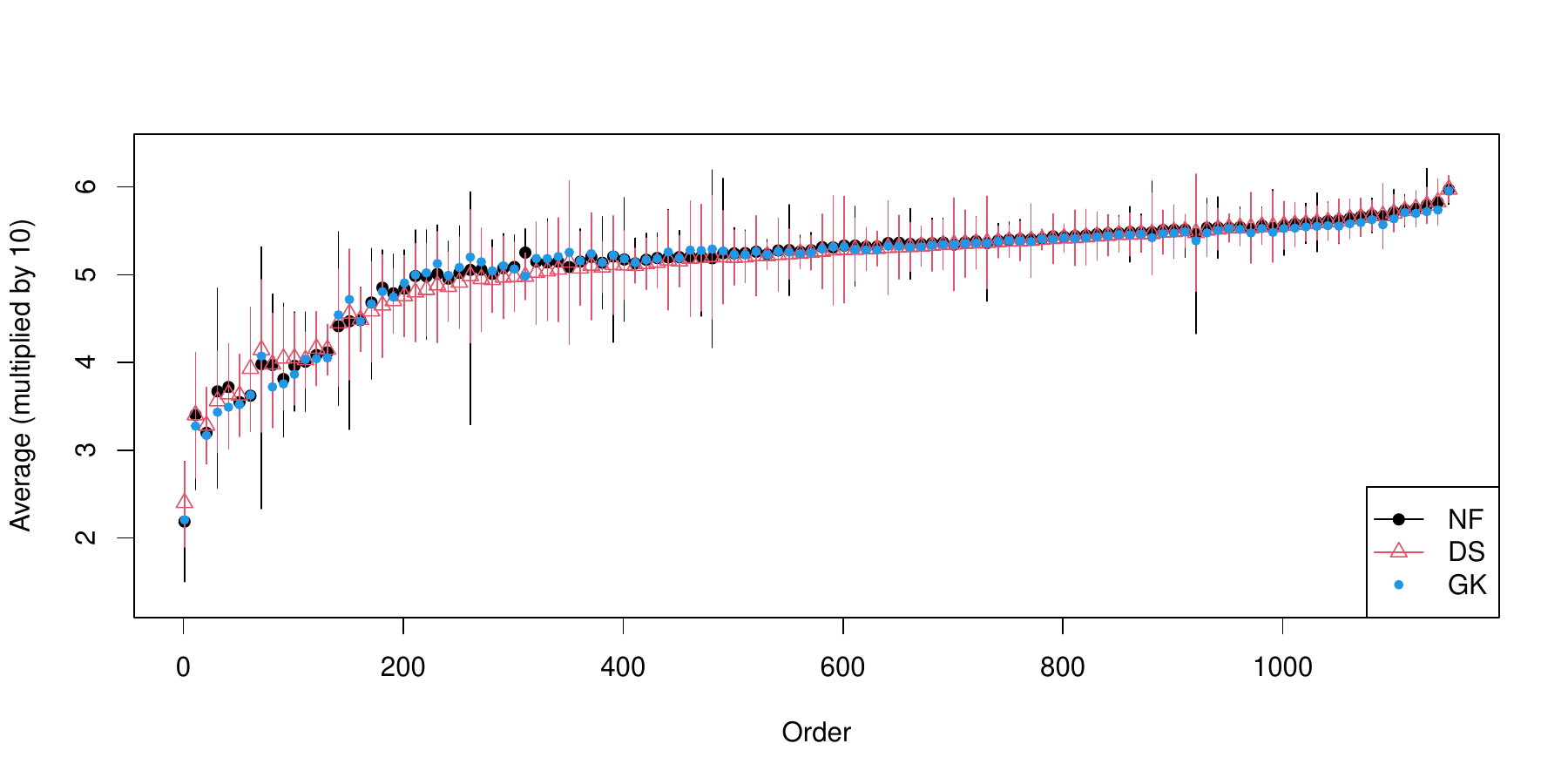}
\caption{ Empirical Bayes estimates with $95\%$ confidence intervals of three methods for selected 120 subjects. } 
\label{fig:app-Est}
\end{figure}

Figure~\ref{fig:app-KDE} displays the estimated conditional density of $(x_i,\log s_i^2)$ given three representative values of the degrees of freedom, corresponding to lower, middle, and upper ranges of $k_i$.  
The results suggest that standard parametric assumptions, such as joint normality, may be inadequate for this dataset.
In Figure~\ref{fig:app-Est}, we present posterior mean estimates of the mean parameter $\theta_i$ obtained by the three methods.  
For DS and NF, $95\%$ confidence intervals are also shown for a subset of 120 players.  
The NF and DS methods tend to produce overly shrunk estimates of $\theta_i$, with the parametric DS method yielding nearly identical estimates across players.  
In contrast, the proposed NF method produces shrinkage estimates similar to DS and GK for players with larger sample averages, while yielding substantially different estimates for players with smaller sample averages.  

%
We next assess the predictive performance of the three methods.  
For each player $i$, the batting records are randomly split into training and evaluation samples to mimic a realistic prediction scenario.  
Specifically, the first $\lfloor n_i/2 \rfloor$ observations are used for training, and the weighted average based on the full dataset for player $i$ is treated as the ground truth.  
We restrict attention to players with more than 1 degrees of freedom in the training sample, resulting in 891 players for the validation study.
Using the training data, we compute $\bar{x}_i$ and $S_i$, and obtain empirical Bayes estimates from the proposed NF method as well as the DS and GK methods.  
We also construct $95\%$ confidence intervals for NF and DS.  
Performance is evaluated in terms of mean squared error (MSE) and empirical coverage probability (CP), computed separately for subgroups defined by thresholds on $\bar{x}_i$.  
This stratification allows us to examine how estimation accuracy varies across players with low, moderate, and high batting averages.

The results are summarized in Table~\ref{tab:app}. 
The proposed NF method exhibits stable performance across all subgroups.  
Its MSE is smaller than the other methods, and its CP stays close to the nominal level in all regimes.  
In contrast, the parametric empirical Bayes methods (DS and GK) display substantial heterogeneity in performance.  
For players with smaller values of $\bar{x}_i$, these methods suffer from severe over-shrinkage, resulting in inflated MSE and small coverage.  
Although their performance improves for larger $\bar{x}_i$, estimation accuracy remains highly sensitive to signal magnitude.
The average interval length (AL) results indicate that NF produces efficient interval estimation, further illustrating the superiority of NF.
This difference is likely attributable to the restrictive parametric assumptions imposed on the prior distribution, which may fail to adequately capture heterogeneity in batting ability.  
Overall, these results highlight the advantage of the proposed NF approach in providing robust and uniformly reliable inference across a wide range of signal strengths.

\begin{table}[htb!]
\centering
\caption{ Mean squared error (MSE) of point estimates, and coverage probability (CP) and average length (AL) of $95\%$ confidence intervals, evaluated for sub-samples separately.}
\label{tab:app}

\medskip
\begin{tabular}{ccccccccccccccc}
\hline
&&&& \multicolumn{3}{c}{MSE ($\times 100$)} && \multicolumn{2}{c}{CP (\%)} && \multicolumn{2}{c}{AL}\\
 &  & \# players &  & NF & DS & GK &  & NF & DS & & NF & DS \\
\hline
$x_i\leq 3.5$ &  & 42 &  & 13.0 & 28.4 & 17.8 &  & 90.5 & 88.1 &  & 1.49 & 1.57 \\
$3.5<x_i\leq 5.0$ &  & 167 &  & 9.33 & 10.4 & 17.3 &  & 91.6 & 96.4 &  & 1.30 & 1.38 \\
$5.0<x_i\leq5.5$ &  & 456 &  & 1.98 & 1.97 & 1.88 &  & 94.7 & 99.6 &  & 0.72 & 0.94 \\
$5.5<x_i$ &  & 226 &  & 1.50 & 1.63 & 1.26 &  & 96.9 & 99.1 &  & 0.59 & 0.76 \\
\hline
Overall &  &  &  & 3.76 & 4.71 & 5.37 &  & 94.5 & 98.3 &  & 0.83 & 1.00 \\
\hline
\end{tabular}
\end{table}

%% file: conclude.tex
This paper develops a new nonparametric empirical Bayes framework for inference with unequal and unknown variances.
By modeling the joint marginal distribution of $(x_i,s_i^2)$ and characterizing the posterior distribution through a moment generating function representation, the proposed approach extends classical $f$-modeling beyond point estimation and enables full posterior inference without specifying a parametric prior.
This addresses a long-standing limitation of traditional empirical Bayes methods, which are often perceived as being restricted to posterior means or variances.

The proposed method accommodates arbitrary dependence between the mean and variance parameters and remains fully nonparametric.
As a result, it provides a unified framework for point estimation, confidence interval construction, and multiple hypothesis testing under heteroscedasticity.
Theoretical analysis establishes regret bounds for the empirical Bayes estimator and error bounds for posterior approximation, while numerical studies demonstrate strong empirical performance in both simulated and real data settings.
In particular, the proposed method achieves efficient shrinkage estimation and reliable uncertainty quantification across a wide range of data-generating scenarios.

More broadly, our results clarify that $f$-modeling, when combined with an appropriate functional representation such as the moment generating function, is capable of recovering the entire posterior distribution.
This insight opens new possibilities for empirical Bayes inference in complex settings where parametric assumptions are difficult to justify.
Future work includes extensions to models with covariates, alternative observation models beyond the normal family, and scalable implementations for ultra-large-scale inference problems.